\documentclass[a4paper,11pt]{article}
\pdfoutput=1 

\usepackage{jcappub} 

\usepackage{float}
\usepackage{color}

\def\beq#1{\begin{equation}\label{#1}}
\def\eeq{\end{equation}}
\def\beqa#1{\begin{eqnarray}\label{#1}}
\def\eeqa{\end{eqnarray}}

\def\mycomment#1{\relax}
\title{\boldmath Electroweak phase transition and entropy release in the early universe}
\def\beq{\begin{equation}}

\author[a]{A. Chaudhuri,}
\author[a,b]{and A. Dolgov}

\affiliation[a]{Novosibirsk State University,  Novosibirsk, 630090, Russia}
\affiliation[b]{ITEP,  Bol. Cheremushkinsaya ul., 25, 117218 Moscow, Russia}

\emailAdd{arnabchaudhuri.7@gmail.com}
\emailAdd{dolgov@fe.infn.it}

\abstract{
It is shown that the vacuum-like energy of the Higgs potential at non-zero temperatures leads, in the course of the 
cosmological expansion, to a small but non-negligible  rise of the entropy density in the comoving volume. This
increase is calculated in the frameworks of the minimal standard model. The result  can have a noticeable effect 
on the outcome of baryo-through-leptogenesis.
}

\begin{document}
\maketitle
\flushbottom

\section{Introduction \label{s-intro}}

As it is well known, the entropy density in the primeval plasma is conserved in the course of the cosmological
expansion, if the plasma is in thermal equilibrium state with negligibly small chemical potentials of all particle
species, see e.g. books~\cite{1,2}. In other words
\beq
s =  \frac{\rho + {\cal P}}{T}{\,a^3} = const ,
\label{s-const}
\eeq
where $T(t)$ is the plasma temperature, $a(t)$ is the cosmological scale factor, and $\rho $ and $\cal{P}$ are respectively
the energy density and the pressure of the plasma. 

Normally the state of matter in the early universe is quite close to the equilibrium because the 
reaction rate, $\Gamma \sim \alpha^n T$, is much faster than the cosmological expansion rate,
$ H = \dot a/a\sim T^2/ m_{Pl} $. Here $n = 1,2$ represents decays and two-body reactions respectively,
$m_{Pl} = 1.2 \cdot 10^{19} $ GeV is the Planck mass and
$\alpha $ is the coupling constant of the particle interactions. Typically $\alpha \sim 10^{-2}$. 
The above estimate for $\Gamma$ is presented for high temperatures, higher than the masses of the participating particles. 
The condition of equilibrium, $\Gamma > H$, is satisfied at the temperatures $T <  \alpha^n m_{Pl}$
up to a constant factor of order unity. Since $m_{Pl}$ is huge, thermal equilibrium existed during most of the
universe history, if  the reaction rate is not anomalously weak, i.e. $\alpha \lll 1$.

In thermal equilibrium, the occupation number (or what is the same, the distribution function)  of any particle species 
is determined by 
two parameters only, the chemical potential, $\mu_j$, of each type of particles and the common temperature of all species. An 
exception is the Bose condensed state, when the chemical potential reaches the maximum value $\mu = m_B$, where $m_B$ is
the boson mass. But still even in this case the system state is also determined by two parameters: the amplitude of the condensate and the temperature of the particles above the condensate. For a system of that kind, entropy,  $s$, surely rises.

In equilibrium the distribution functions with $\mu_j =0$  have the usual  Bose-Einstein or Fermi-Dirac form:
\beq
 f (E) = \frac{1}{\exp (E/T) \pm 1}\,,
 \label{f-eq}
 \eeq
where $E=\sqrt{q^2 + m^2}$ is the particle energy, $q$ is its momentum, and $m$ is the particle mass.

Using this form of the distributions and the expressions:
\beq 
\rho = \int \frac{d^3 q}{(2\pi)^3}\, E f(E),\,\,\, {\cal P} =  \int \frac{d^3 q}{(2\pi)^3}\, \frac{q^2}{3E} f(E),
\label{rho-p}
\eeq
one can easily prove the validity of the conservation law (\ref{s-const}). Moreover, the conservation law (\ref{s-const}) remains true for  any
form of  $ f (E/T) $ with an arbitrary function $T(t)$.

Still there might be several realistic regimes during the universe history when $s$ was not conserved. For example, if  the universe 
was at some epoch dominated by primordial black holes with small masses ~\cite{dnn}, the entropy release could be very high, so that it might essentially delete all
preexisting baryon asymmetry. 
 
A large  amount of entropy  could be produced if the primeval plasma underwent the first order phase 
transition at some early period of the cosmological evolution. Unfortunately, we do not know for sure if such phase
transition(s) indeed took place. A large entropy production might happen, in particular, 
during the QCD phase transition at $T \sim 100-200$ MeV. However
due to strong technical problems  the order of the QCD phase transition in cosmology is not known. For a review
see e.g. ref.~\cite{QCD-pt}

Some, realistic but most probably very weak entropy production, took place during the freeze-out of dark matter (DM)
particles. However, usually the fraction of DM density was quite low at the freezing and the effect is tiny.
 
Possibly the largest entropy release in the standard model took place in the process of the electroweak transition 
from symmetric to asymmetric electroweak phase in the course of the cosmological cooling down. In principle, the 
transition could be either first order or second order, even very smooth crossover. Theoretical calculations say that 
in the minimal standard model with one Higgs field the transition is the mild crossover. However, in an extended 
theory with several higgses, the transition could be even first order with significant supercooling~\cite{EW-first-order,5,6,7,8,3}.

According to the electroweak (EW) theory at the temperatures higher than a critical one, $T > T_c$, 
the expectation value of the Higgs field, $\phi$,
in the plasma is zero and the EW-symmetry is unbroken~\cite{linde}.  When the temperature 
drops below $T_c$, a non-zero  $\langle \phi \rangle $ is created, which  gradually rises, with decreasing
temperature, upto the vacuum
expectation value $\langle \phi \rangle = \eta$, see below. Such a state does not satisfy the conditions necessary for the
entropy conservation and an entropy production is expected.

Here we calculate the entropy release in the course of the transition from the phase with unbroken electroweak symmetry
to the symmetry broken phase.
Presumably in the minimal EW-theory the mild cross-over regime is realized, so we make the 
calculations under this assumption.

\section{Theoretical Framework }

We take the Lagrangian of theory in the following slightly simplified form:
\beq
{\cal L} =  \frac{1}{2} g^{\mu\nu}  \partial_\mu \phi \, \partial_\nu \phi -U_\phi (\phi)
+ \sum_ji \left[ g^{\mu\nu}  \partial_\mu \chi_j^\dagger \, \partial_\nu \chi _j-U_j  (\chi_j)\right] + {\cal L}_{int}
\label{L}
\eeq
where the Lagrangian of the Higgs boson interactions with fields, $\chi_j$, can be taken as: 
\beq
 {\cal L}_{int} =\phi  \sum_j g_{j} \chi_j^\dagger \chi_j .
 \label{L-int}
 \eeq
 The summation is made over all relevant fields $\chi_j$.
 
The equations of motion for the homogeneous classical field $\phi(t) $ and for the operators  of the quantum fields
$\chi _j({\bf x}, t)$ in FRW cosmological backgrounds have the form:
\begin{eqnarray}
&&\ddot \phi + 3H \dot \phi + U'_\phi (\phi) -  \sum_\chi g_{j} \chi_j^\dagger \chi_j = 0,
\label{ddot-phi} \\
&&\ddot \chi_j + 3 H \dot \chi_j - \frac{1}{a^2} \Delta \chi_j  - U'_{j}(\chi_j) - g _j\phi \chi_j = 0,
\label{ddot-chi}
\end{eqnarray}
where $U'_\phi = dU_\phi /d\phi$, $U'_j (\chi_j) =  dU_j/d\chi_j$,
$a(t)$ is the cosmological scale factor, and
$H = \dot a /a $ is the Hubble parameter.

The equation of motion for the classical field $\phi$ (\ref{ddot-phi}) is often taken as:
\beq
\ddot \phi + (3H + \Gamma) \dot \phi + U'_\phi (\phi)  = 0,
\label{ddot-phi} 
\eeq
where $\Gamma$ is the decay width of the $\phi$-boson. Such equation is obtained by thermal averaging of the interaction
term~(\ref{L-int}) and, strictly speaking,  
is valid only for quadratic potential $U_\phi \sim \phi^2$. Generally the equation is more complicated
and may be even non-local in time~\cite{AD-SH}. Still the above equation is sufficiently accurate for an order of magnitude 
estimates.

The self-potential of $\phi$ with the temperature corrections can be written as:
\begin{eqnarray} 
\label{pot-phi}
U_\phi (\phi) = \frac{\lambda}{4}(\phi^2-\eta^2)^2+  \frac{T^2 \phi^2}{2}\,\sum_j \,h_j \left( \frac{m_j(T)}{T} \right) , 
\end{eqnarray}
where according to experiment the vacuum expectation value of $\phi$ is equal to $\eta=246$ GeV and the quartic
self-coupling of $\phi $ is $\lambda=0.13$~\cite{4}. Here $T$ is the plasma temperature and
$m_j(T)$ is the mass of the $\chi_j$-particle at temperature $T$, see below, eqs.~(\ref{phi-min}) and (\ref{m-of-T}).

The last temperature dependent term in eq.~(\ref{pot-phi}) appears as a result of thermal averaging of the interaction (\ref{L-int}). 
It includes the contributions of $\phi$ itself and of all particles $\chi_j$. The summation over $j$ means the summation over all these 
particles. The functions $h_j  (m_j /T) $ are positive and proportional to $g^2_j$.  At high temperatures, $T > m_j (T)$, it is multiplied  
by a constant factor. 
At low temperatures,  $T < m_j(T)$, the function $ h_j(m_j /T)$ is exponentially suppressed, $\sim \exp [-m_j(T)/T]$.

In what follows we will be mostly interested in the contribution of fermions. Their Yukawa coupling constants to the Higgs field are 
determined by their masses at zero temperature, $m_f = g_f \eta$. 
According to the results presented e.g. in ref.~\cite{3} the fermionic loop with single
fermion species gives $h_f (0)= m_f^2/(6\eta^2) $. For quarks this number should be multiplied by three  due to the quark colors.
So e.g. for $t$-quark $h_t (0) = m_t^2/(2\eta^2) =0.25$. 
The masses of all particles depend on the temperature, $m_j = m_j (T)$, because the masses are proportional to the 
expectation value of the Higgs field and the latter is proportional to the temperature dependent value of $\phi$ at the minimum of the 
potential (\ref{pot-phi}):
\beq 
 \phi^2_{min} (T) = \eta^2 - (T^2/\lambda) \sum_j h_j \left(\frac{m_j (T)} {T}\right) 
\label{phi-min}
\eeq
and correspondingly
\beq 
 m^2_f (T) = g^2_f \phi^2_{min} (T) = g^2_f \left[ \eta^2 - (T^2/\lambda) \sum_j h_j \left(\frac{m_j (T)} {T}\right) \right]  .
\label{m-of-T}
\eeq

Here $j=f$ is the index of $\chi_f$-particle which acquires mass through a non-zero expectation value of $\phi$. 
The summation in the r.h.s. of this equation is made over all particles, $\chi_j$ and $\phi$. So for an accurate 
determination of all particle
masses at non-zero $T<T_c$ we need to solve the whole system of equations for all values of $f$. However, because of large
mass differences among the fermions in the standard model only the term with largest $g_f$ (or the highest mass fermion)
contributes to the sum.

As we have mentioned above, at the temperatures higher than the critical value $T_c$ the expectation value of the Higgs field vanishes, while at $T<T_c$
the expectation value $\langle \phi \rangle $ becomes non-zero and all particles acquire non-zero, temperature 
dependent masses $m_f (T)$. As we have already mentioned, a particle gives noticeable contribution to thermal
mass of the Higgs field if  $m_f(T)\lesssim T$. 
According to eq. (\ref{m-of-T}), the critical temperature is determined by the equation:
\beq
T_c^2 = \frac{\lambda \eta^2}{  \sum h_j (0)} .
\label{T-c}
\eeq
In terms of $T_c$ the value of $\phi$ at the minimum of the potential or, what is the same, the expectation value of $\phi$ 
in the plasma can be written as
\beq 
\phi^2_{min} = \eta^2 \left[ 1 - \frac{T^2}{T^2_c}\,\frac{ \sum h_j (m_j/T)}{\sum h_j (0)}  \right] 
= \eta^2 \left[ 1 - \frac{T^2}{T^2_c}\,\frac{ h_{tot}(m/T)}{ h_{tot}(0)}  \right] ,
\label{phi-min}
\eeq
where to simplify the equations we introduced the notations
\beq 
 \sum h_j (m_j/T) \equiv h_{tot} (m),\,\,\,  \sum h_j (0) \equiv h_{tot} (0) .
 \label{h-tot}
 \eeq

To describe the behavior of $h_{tot} (m)$ we need to define for each particle (fermion) the temperature at which its 
temperature-dependent mass becomes equal to the temperature,  $m_f (T) =T$.  Above this temperature the quark
contribution to $h_q (m_q(T)/T) $ is equal to $h_q=g_q^2/2=m_f^2 (0) /(2\eta^2)$ (later on we use the notation 
$m_f(0) \equiv m_f$).
The lepton contribution is  $ h_l = m_l^2 /(6\eta^2)$.
Below this temperature, $h_f$ is exponentially suppressed and can 
be neglected. Since masses of the quarks and leptons are very much different (except for $u$ and $d$ quarks) we may approximate
$h_{tot} (m)$ as a succession of theta-functions dominated by a single fermion $f$ in the temperature range
$T_{min}^{(f')} \leq T \leq T_{min}^{(f)}$, where $f'$ is the heavier fermion nearest by mass to $f$.

According to eq.  (\ref{m-of-T}) and (\ref{T-c}),  $m_f(T)$ 
would remain smaller than $T$ for the temperatures higher than
\beq 
(T^f_{min})^2 = \frac{g_f^2 \eta^2}{ 1 + \left(g_f^2 \eta^2/T_c^2\right) \left[h_{tot} (m) /h_{tot} (0) \right] }.
\label{T-min} 
\eeq
As we have mentioned above, $h_{tot}(m)$ is dominated by  the single contribution of the fermion $f$
at the temperatures near $T^f_{min}$. So eq.~(\ref{T-min}) is reduced to 
\beq 
(T^f_{min})^2 \approx \frac{g_f^2 \eta^2}{ 1 + (N_f/3)\left(g_f^4 \eta^2/g^2_tT_c^2\right) },
\label{T-min-simpl} 
\eeq
where $N_f = 3$ for quarks and $N_f = 1$ for leptons. Note, that
only for t-quarks two terms in the denominator are comparable, while for lighter quarks and leptons $T_{min}^f \approx m_f$. 
At the temperatures in the interval $ T_{min}^{(t)} \leq T \leq T_c$ $t$-quark dominates, while   
at $T_{mn}^b<T<T_{min}^t$, $b$-quark gives the dominant contribution, and so on.

The potential $ U_\phi(\phi) $ is chosen in such a way that it vanishes when $\phi$ takes its vacuum expectation value, 
$\phi = \eta$. 
It ensures zero vacuum energy of the classical field $\phi$. 
For nonzero temperature, $\phi < \eta$ and $U(\phi_{min} ) \neq 0$:
\beq
U_\phi (\phi_{min}) = 
\frac{h_{tot} (m) T^2 \eta^2}{2} 
\left[ 1 - \frac{T^2  h_{tot}(m)  } {2\lambda \eta^2 }\right] = 
\frac{h_{tot} (m)T^2 \eta^2}{2} \left[ 1 - \frac{T^2}{2 T^2_c}\, \frac{ h_{tot}(m)}{ h_{tot} (0)} \right].
\label{U-of-phi-min}
\eeq

Let us note that the equations presented above are true in the broken phase, when there are one real Higgs field and
massive three component intermediate W and Z bosons.

To calculate the entropy density we need the expression for the energy-momentum tensor:
\begin{eqnarray}
T_{\mu\nu} &=& \partial_\mu \phi \, \partial_\nu \phi - 
g_{\mu\nu} \left( g^{\alpha\beta} \partial_\alpha\phi \, \partial_\beta \phi - U_\phi (\phi) \right) 
\\ \nonumber 
&+& \sum_j \left[ \partial_\mu\chi_j^\dagger \, \partial_\nu\chi_j + \partial_\nu\chi_{j}^\dagger \, \partial_\mu \chi_{j} -
g_{\mu\nu} \left( g^{\alpha\beta} \partial_\alpha\chi^\dagger_{j} \, \partial_\beta\chi_{j} - U_j (\chi_j) + 2 {\cal L}_{int}  \right) \right] ,
\label{T-mu-nu}
\end{eqnarray}
where 
${\cal L}_{int} $ is given by eq. (\ref{L-int}).

The operators of the energy density and pressure density for homogeneous classical field $\phi$ and all other fields $\chi_j$ (quanta of $\phi$
should be  included there) in the Friedmann-Robertson-Walker background have the form:
\begin{eqnarray}
\rho &=& \dot\phi^2/2 + U_\phi (\phi)  + 
\sum_j \left[ \dot \chi_j^\dagger \dot \chi_j + \partial_l\chi_{j}^\dagger\, \partial_l\chi_{j }/a^2  + U_j (\chi_j) \right] - {\cal L}_{int}; \label{rho} \\
{\cal P} &=& \dot\phi^2/2 - U_\phi (\phi)  + 
\sum_j \left[ \dot \chi_j^\dagger \dot \chi_j - (1/3)\partial_l\chi_{j}^\dagger\, \partial_l \chi_{j} /a^2  - U_j (\chi_j) \right] + {\cal L}_{int},
\label{p} 
\end{eqnarray}
where $\partial_i \chi$ is the space derivative.

The sum $(\rho +{\cal P})$ enters into the expression for the entropy density and into the equation governing 
the evolution of $\rho$, see below eq.~(\ref{dot-rho}). It is equal to:  
\beq
\rho + {\cal P} =  \dot\phi^2 + \sum_j \left[ \dot \chi_j^\dagger \dot \chi_j + 
\frac{2}{3 a^2}\, \partial_l\chi_{j}^\dagger \, \partial_l\chi_{j }   \right] .
\label{rho-plus-P}
\eeq

It can be verified that $\rho(t)$ indeed satisfies the covariant conservation law: 
\beq
\dot\rho = -3 H (\rho + {\cal P} ),
\label{dot-rho}
\eeq
if we use the equations of motion (\ref{ddot-phi}, \ref{ddot-chi}) and neglect the terms containing total spatial divergence.

Let us calculate now the variation of entropy in the course of the cosmological expansion, 
using the definition (\ref{s-const}), expressions (\ref{rho}, \ref{rho-plus-P}), and the equations
of motion (\ref{ddot-phi}, \ref{ddot-chi}). The calculations will be greatly simplified if we assume that the energy density
consists of two parts, the energy density of the field $\phi (t)$ sitting at the minimum of the potential 
and of relativistic matter, so the expression for $\rho$ becomes:
\beq
\rho \approx U_\phi (\phi_{min} ) + \frac{\dot \phi^2}{2}  + \frac{\pi^2 g_*}{30} T^4,
\label{rho-approx}
\eeq
and
\beq
\rho + {\cal P} \approx \dot \phi^2 + \frac{4}{3}\,\frac{\pi^2 g_*}{30} T^4,
\label{rho-plus-P-approx}
\eeq
where $g_*\sim 10^2$ is the effective number of particle species at or near the electroweak phase transition. It is a  
function of temperature,
decreasing in the course of the cosmological cooling down.
Equation (\ref{rho-approx}) is valid in the limit of instant thermalization.

The oscillations of $\phi$ around $\phi_{min} $ are quickly damped, so we take $\dot \phi = \dot \phi_{min}$ and neglect $\dot \phi^2$ 
in what follows, because the evolution of $\phi_{min}$ is induced by the universe expansion which is quite slow.
In this approximation we obtain the single differential equation governing the temperature evolution with time, or what is
more convenient, with the scale factor. Under these assumptions equation (\ref{dot-rho}) can be rewritten as
\beq
\frac{\dot T}{T} \left[ 
h_{tot} (m) \eta^2  T^2 \left( 1  - \frac{T^2}{T^2_c}  \frac{ h_{tot} (m) }{ h_{tot}(0) } \right)  + \frac{4 \pi^2 g_*}{30} T^4\right]  
= - 4H \frac{\pi^2 g_*}{30} T^4 .
\label{dot-T}
\eeq
Here equations (\ref{U-of-phi-min}), (\ref{rho-approx}), and (\ref{rho-plus-P-approx}) are used and the time derivative 
of $h_{tot}(m)$ is neglected, because $h_{tot}$ is supposed to be the succession of the step functions. 

Let us note that the equation (\ref{dot-T}) does not take into account the modification of the temperature evolution due to
annihilation of non-relativistic species, as e.g., the well known heating of plasma by $e^+e^-$-annihilation, which takes place
at $T$ below $m_e$. We disregard this effect because, if the annihilating particles are in thermal equilibrium state with vanishing
chemical potential, the entropy density in this process is conserved. 

It is convenient to introduce the parameters:
\beq
\kappa = \frac{30 h_{tot}(m)}{4 \pi^2 g_*}, \,\,\,\, \nu  = \frac{\kappa \eta ^2}{ 2 T_c^2},
\label{kappa-nu}
\eeq 
so equation (\ref{dot-T}) turns into
\beq 
 \frac{\dot a }{a} = - \frac{\dot T}{T} \left[ \frac{\kappa \eta^2}{T^2} \left( 1 - \frac{T^2}{T_c^2}  \frac{ h_{tot} (m) }{ h_{tot}(0) } \right) 
 + 1 \right] .
 \label{dot-a}
 \eeq
 In the case when the heaviest particle mass, that of $t$-quark, is lower than the temperature, we can take $h(m) = h (0)$ 
 and equation (\ref{dot-a}) can be easily integrated resulting in:
 \beq 
 \frac{ a(T) T}{a_c T_c} = x^{2\nu} \exp\left[ \nu \left(\frac{1}{x^2} -1 \right)\right] ,
 \label{a-Tc}
 \eeq
 where $x = T/T_c$ and the cosmological scale factor $a_c$ is taken at $T=T_c$.

Taking $T_c$ from eq.~(\ref{T-c}), we find for $t$-quark:
 \beq 
 (x_{min}^t)^2\equiv \left(\frac{T_{min}^t}{T_c} \right)^2 = \frac{g_t^2 (\eta/T_c)^2}{1 + g_t^2 (\eta/T_c)^2} = 
 \frac{ m_t^2  /T_c^2}{ 1 + m_t^2/T_c^2 } .
 \label{x-min}
 \eeq
 
Since $t$-quark is the heaviest among the standard model particles, its contribution to the entropy release is the largest
at high temperatures. But it quickly disappears when temperature drops below the running mass of $t$-quark, i.e. at  
$T<m_t (T)$.  On the other hand, lighter particles become efficient
at lower $T$. Due to that, their contribution remain more or less the same as that of $t$-quark. 
 Moreover, as one can see from eqs. (\ref{kappa-nu}), the  effect is inversely proportional to the number of the particle
 species $g_* (T)$. It drops down from $g_* = 106.75$ at the electroweak phase transition to 10.75 at 
 the temperatures below the muon mass.  So, as we see in the next section, it considerably amplify the contribution of
 light leptons into the entropy increase in the course of the cosmological expansion.
 
 \section{Calculations and Results}

We start with the calculations of the contribution to the entropy from the heaviest particles.
To this end we need the numerical values of their coupling constants with the Higgs boson. According to 
the experimental data, they are 
$g_t^2 = 0.25$, $m_t =173 $ GeV,
$\lambda = 0.13$, $g_W^ 2 = 0.13$, $g_Z =  0.1$. The Yukawa coupling constants of lighter 
fermions scale as the ratio of the masses, $g_f = g_t (m_f / m_t)$. 

With  the account of $t$-quark only the critical temperature, according to eq.~(\ref{T-c}) is $T_c^2/\eta^2 =   2\lambda \eta^2/m_t^2  \approx 0.53$

The contribution of other heavy particle makes this ratio twice smaller:
\beq
\frac{T_c^2}{\eta^2} =   \frac{ 2\lambda \eta^2}{m_t^2+m_w^2+m_z^2+m_H^2}  \approx .25 
\label{Tc-ovet-eta}
\eeq


Let us estimate now the values of parameters $\kappa$ and $\nu$ of eq.~(\ref{kappa-nu}). As we mentioned above,
$h_{tot} (m)$ is a collection of theta-functions dominated by single contribution of a fermion with mass specified below
eq.~(\ref{T-min}). Correspondingly at $T> T^f_{min} \approx m_f$ the contribution of  a fermion to $h_{tot} (m)$ is equal to
$h_f = N_f  m^2 /(6\eta^2)$, where $N_q = 3$ and $N_l = 1$. Hence the contribution of all fermions lighter than $t$-quark is
\beq
\kappa = \frac{5}{4\pi^2} \sum_f  \left[\frac{N_f  m^2_f }{ g_*( T_{min}^f ) \eta^2} \right],
\label{kappa-sum}
\eeq
where  $g_*( T_{min}^f )$ is the number of relativistic particle species present in the plasma at $T \sim T_{min}^f$. So 
$\kappa$ and $\nu \approx \kappa$ are both small numbers, $\nu \approx 0.007$ for $g_* =106.76$ and 
$\nu \sim 0.07$ for $g_* = 10.75$. One should keep in mind, however, that these small numbers are mulitipleid by
$ (T_c /T_{min}^2)^2 \sim T_c^2/m_f^2 $, which can be very large. It is interesting that the product $\nu (T_c/m_f)^2$ 
essentially does not depend upon the mass and the effect is the larger for smaller masses due to a decrease of $g_*$.

Using equation (\ref{a-Tc}), we find that the relative increase of the entropy is
\beq 
\frac{\delta s}{s} = \sum_f x_{j,min}^{6\nu_f} \exp\left[ 3\nu_f \left(\frac{1}{x_{f,min}^2} -1 \right)\right]  - 1 ,
\label{delta-s}
\eeq
where $\nu_f$ includes only contribution from single fermion $f$ and, according to eq.~(\ref{T-min-simpl}), 
\beq
x_{f,min} = \left(\frac{T^f_{min}} {T_c}\right)^2 \approx \frac{g_f^2 \eta^2/T_c^2}{ 1 + (N_f/3)\left(g_f^4 \eta^2/g^2_tT_c^2\right) },
\label{x-f-min}
\eeq
With an exception for $t$-quark this expression is reduced to a very simple one 
\beq
x_{f,min} = (m_f/T_c)^2,
\label{x-min-not-t}
\eeq
and correspondingly 
\beq
\frac{3\nu_f}{x_{f, min}^2} = \frac{15 N_f}{4\pi^2 g_*}\,\left( \frac{T_c}{\eta}\right)^2 
\label{3nu-x}
\eeq

For example, the electron contribution to the relative rise of the entropy is $(\delta s/s)_e = 1.8 \%$.  At  temperatures below the muon
mass, $g_* =14.25$ and thus the muon contribution is $(\delta s/s)_\mu = 1.3 \%$. The contribution of $\tau$-lepton is 
$(\delta s/s)_\tau = 0.25 \%$, because at $T=180$ GeV, $g_* = 75.75$. The contribution of three quark families in this temperature
range is 12 times larger and brings about $3\%$. The contribution from $t$-quark is $(\delta s/s)_t \approx 1\%$. The contribution from $b$-quark is almost the same but it is to be noted that the contribution came from a bigger range of temperature ($T_{min}$ for $b$-quark is $\approx 4 GeV$).  Moreover, the contributions of the  lighter $s$, $u$, and $d$ quarks are slightly 
enhanced because they remain alive down to the QCD phase transition at about 150 MeV, when $g_*$ is 72.25. The contribution of Higgs boson $(\delta s/s)_H = 1 \%$ and that of Gauge bosons is $(\delta s/s)_{W,Z} \approx 2 \%$.

We need to take into account that neutrinos are decoupled from the electromagnetic component of the cosmic plasma at
$T_{e} \approx 1.9 $ MeV for $\nu_e$ and at $T_{\mu,\tau} \approx 3.1 $ MeV for $\nu_\mu$ and $\nu_\tau$,
see e.g.~\cite{nu-rev}.  This effect would lead to the decrease of the effective number of species from 10.75 
to $g_* = 5.5$ and the rise of the electron contribution up to $(\delta s/s)_e =3.6\%$.

\begin{figure}[h!]
\includegraphics[width=.8\textwidth]{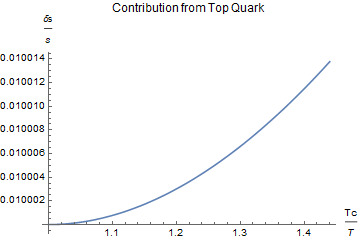}  
  \caption{ }
\label{fig1}  
\end{figure}

As an illustration, the entropy production from $t-$quark as a function of temperature is presented in Fig. $\ref{fig1}$. Contributions to the entropy release comes till $T_{min}$ of every particles.

Similar calculations have been done for other particles and the entropy release has been calculated in the huge range of temperature, starting from the temperature of the Electroweak phase transition down to the mass of electron ($511KeV$). So, in the range from $GeV$ to $keV$ scale, we find that the total amount of entropy is increased by about \textbf{$13 \%$}.

\section{Discussion and Conclusion} 
It is shown that the total entropy release in the course of the electroweak symmmetry breaking is quite noticible even in the frameworks of minimal Standard model of particle physics. We have assumed here that Electroweak phase transition is second order(or smooth/mild crossover). It is to be noted that $g_*$ decreases as the temperature falls down. But as we go to very low temperature scale, the minimum temperature ($T_{min}$) takes the value of the particle mass and hence we find that the contribution of lighter particles in the process of entropy release is nearly similar to that of the heavy particles, like $t$-quark.

In extended versions of the electroweak theory (e.g., with several Higgs fields) the entropy release may be considerably larger.

\section{Acknowledgements}
Our work  was supported by the RSF Grant N 16-12-10037.

\end{document}